**Nanosheet-stabilized emulsions: ultra-low loading segregated networks and surface energy determination of pristine few-layer 2D materials**

*Sean P. Ogilvie[a]\*, Matthew J. Large[a], Adam J. Cass[a], Aline Amorim Graf[a], Anne C. Sehnal[a], Marcus A. O'Mara[a], Peter J. Lynch[a], Jonathan P. Salvage[b], Marco Alfonso[c], Philippe Poulin[c], Alice A. K. King[a] and Alan B. Dalton[a]\**

[a] University of Sussex, Brighton, BN1 9RH, United Kingdom
[b] University of Brighton, Brighton, BN1 4HP, United Kingdom
[c] Centre de Recherche Paul Pascal – CNRS, University of Bordeaux, 33600 Pessac, France

E-mail: s.ogilvie@sussex.ac.uk; a.b.dalton@sussex.ac.uk

A framework is developed to allow emulsification to be used to fabricate functional structures from, and study the properties of, pristine layered nanosheets. Liquid-exfoliated few-layer graphene and MoS$_2$ are demonstrated to stablize emulsions which exhibit system-scale electrical conductivity at ultra-low nanosheet volume fractions. When deposited on a substrate, the controlled drying dynamics of these emulsions facilitates their application as inks where the lack of any coffee ring effect allows manual deposition of high conductivity films. In order to broaden the range of compositions and subsequently applications, an understanding of emulsion stability and orientation in terms of surface energy of the three phases is developed. Importantly, this model facilitates determination of the surface energies of the nanosheets themselves and subsequently allows design of emulsions. Finally, emulsification by surfactant-exfoliated nanosheets and emulsion inversion using basic solution are demonstrated to allow water-based processing where composition and orientation can be tailored to enable applications.

**Introduction**

Liquid phase exfoliation of pristine graphene and related two-dimensional (2D) materials has enabled assembly of solution-processed nanosheet networks with a broad range of electronic, electrochemical, thermal and mechanical properties.[1] While thin film applications such as printed electronics[2] and energy storage materials[3,4] are well-developed, macroscopic structures are typically limited to random networks in polymer matrices.[5] This limits the range of accessible applications and often requires high loadings to achieve the desired properties. As such, more controlled assembly techniques are required to broaden the range of possible structures and realize enhanced functional properties at low loading level.

Pickering emulsification is a versatile technique for assembling solid particles at the interface between two immiscible liquids, resulting in solid-stablized droplets of one phase in the other[6]. Clearly, for 2D nanosheets, this presents a route towards assembly of structures where the degree of exfoliation is maintained in situ, preserving high number densities of nanosheets, which act as both emulsion stablizer and functional filler. In addition, their atomically-thin nature and correspondingly high specific surface area could allow stabilisation of microscale droplets with nanoscale film thickness, potentially enabling macroscopic functionality at low nanosheet loadings. This is analogous to the segregated network approach demonstrated for carbon nanotube[7] and graphene composites[8], recently extended to enable state-of-the-art battery electrodes[9].

While Pickering emulsification has been studied for clays[10,11], graphene oxide (GO)[12], reduced GO[13] and graphitic multilayers[14–16], emulsions stablized by pristine few-layer nanosheets have not yet been realized. This is likely because of the difficulty in exfoliating these materials in appropriate liquids to allow emulsification. Here, we develop a framework for understanding and design of emulsion stablized by pristine few-layer nanosheets to enable their applications including ultra-low loading functional composites and energy storage materials.

**Exfoliation and emulsification**

The mechanism of Pickering emulsification is that two immiscible liquids partially wet the solid stablizer such that the total interfacial energy is less than that of the oil-water interface.[6,16] This is typically achieved with a high surface tension "water" phase, most often water, and a low surface tension "oil" phase which can be any water-immiscible organic. Given the poor dispersability of pristine 2D materials in water (without surfactant, which acts to stablize the emulsion itself), the most obvious route to formation of these emulsions is exfoliation into the oil phase followed by emulsification with water as shown in Figure 1a, to produce emulsions whose orientation (whether o/w or w/o) is determined by the relative interfacial energies at the three-phase boundary shown in Figure 1b.

In order to realize emulsions stablized by few-layer nanosheets, they must be well-exfoliated before emulsification. This requires solvents which are well-matched in surface energy and Hansen parameters to the nanosheets[17], which precludes water-immiscible organics such as chloroform, ethyl acetate and common monomers such as methyl methacrylate, and also immiscible with water, which precludes common exfoliating solvent such as *N*-methyl-2-pyrrolidone, dimethylformamide, acetone and most alcohols. Using this solvent selection approach, illustrated in Figure 1c, cyclopentanone (CPO) and cyclohexanone (CHO) were identified as water-immiscible solvents for direct exfoliation and emulsification, which also have relatively low boiling point to facilitate subsequent evaporation. In practise, we find higher concentrations and stability for graphene/CHO and $MoS_2$/CPO and hereafter use these as standard exfoliating solvents for these materials.

Nanosheet-stablized emulsions of water droplets in a continuous oil phase can be formed by addition of water to these cycloketone dispersions of few-layer nanosheets followed by simply shaking by hand. These droplets, shown in Figure 1d and 1e, are typically between 10 and 500 µm in diameter and are optically semi-transparent, indicating that the nanosheets form disordered films of <20 monolayers, confirming stabilisation of emulsion by few-layer nanosheets.

**Conductive segregated networks**

These nanosheet-stablized emulsions droplets represent potential building blocks of segregated networks where the templated self-assembly ensures system-scale conductivity with all nanosheets contributing to the network conductivity. As such, the relationship between droplet size and nanosheet volume fraction will inform the resultant properties of the conductive networks.

To characterise this the water-in-cycloketone emulsions were formed with fixed ratio of liquids but varying nanosheet volume fraction and average droplet diameter measured by statistical optical microscopy. Figure 2a shows the average droplet diameter as a function of volume fraction with values between 10 and 500 µm for nanosheet loadings across three order of magnitude below 1 vol.%.

This can be understood in terms of a simple geometric relation equating the surface area of the nanosheets to the surface area of the droplets to give an expression relating droplet diameter and nanosheet volume fraction

$$\langle d \rangle = \frac{6 c_{2D} \langle N \rangle}{\phi} \tag{1}$$

where $\phi$ is the volume fraction of the nanosheets relative to the droplet phase and $\langle N \rangle$ here denotes the area-averaged film interfacial thickness as a number of monolayers, rather than the thickness of the individual nanosheets, and $c_{2D}$ is the interlayer spacing in the bulk material (full derivation in Supporting Information).

It is worth noting that this model assumes constant interfacial film thickness in order for the $\phi^{-1}$ scaling to be realized. In practise, free exponent fitting of the data gives a value of $\phi^{-0.5}$ rather than $\phi^{-1}$, for both graphene and $MoS_2$, as shown in Figure 2a. This is equivalent to the interfacial film thickness increasing with loading level as $\langle N \rangle \sim \phi^{0.5}$, as shown in Figure 2b, consistent with the interpretation that the droplets are being overcoated to multiple nanosheets' thickness. This likely arises from droplets of one liquid in another being formed at smaller droplet size than can be stablized, due to unavailability of nanosheets and/or nanosheet rigidity preventing stabilising of submicron droplets. The inferred interfacial film thicknesses take values between 5 and 50 for graphene and 0.3 and 5 for MoS$_2$, perhaps suggesting some material or solvent influence of interfacial film formation. Nevertheless, the robustness of the $\phi^{-0.5}$ scaling across both materials suggests some well-defined physics governs droplet formation and allow realisation of emulsions stablized by pristine few-layer nanosheets.

It is intuitive that network conductivity will increase with both reduction of droplet size, due to increased parallelisation of the network, and increasing nanosheet volume fraction. Interestingly, these interfacial films do not exhibit percolative behaviour typically associated with nanocomposites; there is no clear percolation threshold because reducing the volume fraction simply increases the droplet size until there is a single large droplet whose diameter is approaching the size of the container. It is interesting to note that emulsions are essentially films in low loading level limit and random composites in the high loading level limit. As such, their conductivity-volume fraction relationship can be fitted to power law scalings, as shown in Figure 2c, which are simply percolation curves, accounting for the scaling of paths in the network, but with an ultra-low near-negligible threshold. These networks have conductivities approaching those of typical graphene-polymer composites[5] and, to the best of our knowledge, are the lowest loading levels ever reported for graphene-based conductive networks, as shown in Figure 2d.

**Emulsion inks**

The formation of disordered interfacial films with controllable thickness presents the possibility of dispersing nanosheets at high concentration with energetic, rather than solely kinetic, stability, highlighting their suitability as inks for deposition of thin films. As illustrated in Figure 1D, the ideal combination of properties for an emulsion ink are realized in water-in-cycloketone emulsions. In addition, nanosheet-coated water droplet sediment onto and are stable in contact with hydrophobic polymeric substrates such as PET, in contrast to on glass where they wet and spread or buoyant oil droplets with rise and burst at the air interface.

This stability of deposited water droplets on polymeric substrates confers a degree of spatial control to deposition of emulsion inks even for drop-wise manual deposition. As shown in Figure 3a, water droplets are stable on substrate until spreading and evaporation of the capping layer of solvent. The exposed graphene-coated water droplet then forms an unstable three-phase interface with the air (only stable for air-in-water), resulting in deformation, drying and collapse of the droplet onto the substrate. By contrast with dispersions, where wetting of the substrate by the liquid results in loss of any spatial control and drying results in some degree of coffee-ring effect, the collapse of these droplets appears to minimise this effect in emulsion inks as shown in Figure 3b.

This uniform drying and spatial control of emulsions, along with the ability to prepare at higher concentrations than dispersions, facilitates drop-wise deposition of nanosheet networks by hand with greater control than drop casting or spray coating from dispersions. The deposited nanosheets form dense packed networks with macroscopic electrical conductivity as shown in Figure 3c. Interestingly, the conductivities exhibit thickness-dependent scaling as observed previously but the macroscopic non-uniformity introduced by manual depositing result in critical thicknesses of ~1 μm, *cf.* 50-200 nm in previous studies of vacuum filtration or inkjet printing.[18,19] Nevertheless, the measured conductivities of graphene films reach bulk-like thickness-independent values of ~3000 S/m as shown in Figure 3c. This can be attributed to the formation of dense packed networks of nanosheets illustrated by the inset SEM image in Figure 3c and AFM height image in Figure 3d. In addition, Raman mapping of individual deposited droplets shows spectra characteristic of liquid-exfoliated few-layer graphene with 2D/G ratio associated with layer number uniform across the droplet (Figure 3e). Furthermore, the G peak intensity is similarly uniform as shown in Figure 3f, indicating that individual emulsion droplets can be deposited without coffee ring effects which may enable inkjet printing of emulsion droplets. Indeed, these emulsions exhibit the expected non-Newtonian properties; they are found to be shear thinning with viscosity given by characteristic power law of the form $\eta \approx 0.1\, \dot{\gamma}^{-0.58}$ (as shown in Supporting Information Figure S2), highlighting the possibility of tuning emulsion droplet size, oil-to-water ratio and composition to give the required viscosity and shear rate combination to allow inkjet printing of individual emulsion droplets.

**Nanosheet surface energy**

In order to realize the full range of applications envisaged, it will be necessary to form nanosheet-stablized emulsions with liquids other than water and cycloketones. However, for the reasons illustrated in Figure 1d, it is quite challenging to use alternative solvents while retaining the high degree of exfoliation required for ultra-low loading applications. In practice, this can be achieved using a solvent transfer step based on liquid cascade centrifugation.[20] Dispersions are prepared in cycloketones as normal and subjected to further centrifugation to sediment the majority of the nanosheets, the supernatant is discarded and the sediment is redispersed into an alternative solvent of choice before immediate emulsification. This allows for production of well-exfoliated materials in solvents where this would not be possible by direct exfoliation such that few-layer nanosheet-stablized emulsions can be produced with relatively arbitrary oil and water phases.

This approach allows us to investigate emulsification of liquids with different surface tensions to modify the three-phase boundary shown in Figure 1b. Having established that graphene, $MoS_2$ and boron nitride (BN) are capable of stabilising water-in-cycloketone emulsions, suggesting preferential wetting of the nanosheets by the cycloketone compared with the water, it was noted that less polar oil and/or water phases would be required to produce oil-in-water emulsions.

The stability and orientation (whether oil-in-water or water-in-oil) of these emulsions is determined by the three-phase boundary and associated interfacial energies and spreading coefficients. These are defined as

$$S_{so} = \gamma_{so} - \gamma_{sw} - \gamma_{ow} \tag{2}$$

$$S_{sw} = \gamma_{sw} - \gamma_{so} - \gamma_{ow} \tag{3}$$

where $S_{so}$ and $S_{sw}$ are the spreading coefficients for solid/oil and solid/water interfaces respectively and the subscripts of the surface energies denote the contributions as shown in Figure 1c. The criterion is typically that they must both have the same sign (positive or negative) for an emulsion to be stable, where one phase preferentially wets the solid stablizer and therefore forms the continuous phase while the other forms the droplet phase, as illustrated in Figure 4a.[6]

From the definitions of the spreading coefficients, and in line with intuition, it can be shown (see Supporting Information) that phase inversion occurs at

$$\gamma_{so} = \gamma_{sw} \tag{4}$$

where the phase which has the lowest interfacial tension with the solid will form the droplets, independently of the interfacial tension of the two phases.

While interfacial tensions between liquids can be measured, it would be preferable to understand the spreading coefficients as a continuous function of the individual and well-known surface tensions of the liquids. To facilitate this, well-established simple models for interfacial tension as a function of surface tension[16,21,22] can be employed, such as the following geometric mean model;

$$\gamma_{ab} = \gamma_a + \gamma_b - 2\sqrt{\gamma_a \gamma_b} \tag{5}$$

For graphene and related materials, as solids, the surface entropy (and therefore surface tension) is poorly-defined and therefore it is more correct to infer the surface *energy* from its interaction with liquids of known surface energy[23] or by inverse gas chromatography.[24,25] As such, liquid-exfoliated graphene is understood to have a surface energy close to 70 mJ/m² based on good exfoliation and dispersion into solvents with surface tensions close to 40 mN/m.

As such, this inversion threshold can be further simplified, by substituting equation (5) into equation (4) to be given in terms of surface energies

$$\sqrt{\gamma_o} + \sqrt{\gamma_w} = 2\sqrt{\gamma_s} \tag{6}$$

where lower surface energies of the liquid phases give o/w and higher surface energies give w/o, and the threshold itself is determined by the surface energy of the solid stablizer; in this case, the layered nanosheets.

In practice, this equation describes all experimental observations in terms of stability and orientation for all combinations of liquids, air and substrate interfaces and nanosheet type (graphene, $MoS_2$ and BN), confirming the nanosheet surface energies to be close to 70 mJ/m$^2$ as shown in Figure 4b. Importantly, this equation only describes all experimental results when considering surface energies (rather than tensions) as the interfacial properties are non-linearly related to individual surface properties. In addition, the same emulsion orientations are observed for graphene, $MoS_2$ and BN suggesting they have little difference in their effective surface energies.

Importantly, using this equation, it is possible to measure the surface energy of layered nanosheets based on inversion of an emulsion by changing its composition. To perform this measurement on well-exfoliated few-layer nanosheets, cycloketone dispersions were diluted with pentane and immediately emulsified with water to determine their orientation as a function of pentane volume fraction. The surface tension of the cycloketone/pentane dispersions was measured and used to calculate bounds of the surface energy of the nanosheets based on the emulsion orientation. Inversion of these emulsions was observed to occur at a pentane volume fraction between 0.90 and 0.95, with a surface tension of ~17 mN/m, as shown in Figure 4c, corresponding to a nanosheet surface energy of 71 ± 0.5 mJ/m$^2$.

These measurements illustrate that the formation of oil-in-water emulsions requires the use of either a very low surface energy oil phase and/or water phase with lower surface energy than water such as ethylene glycol. As a result, oil and water phases which yield oil-in-water emulsions tend to be poor solvents for dispersion of liquid-exfoliated nanosheets. Consequently, an alternative approach is required to allow few-layer nanosheets to be emulsified with liquid phases chosen for the given application. In practice, this can be achieved using surfactant-exfoliated nanosheets where a sufficient amount of the free surfactant has been removed by, for instance, liquid cascade centrifugation followed by redispersion into pristine deionised water, such that only surfactant bound to the nanosheets remains. This allows formation of stable emulsions and provides a route to using water as a universal carrier for nanosheet-stablized emulsions. In addition, by adding ethylene glycol to the water phase, the surface energy can be determined for a range of oil phases and fitted to yield the surface energy of the surfactant-exfoliated graphene (Figure 4d), which is found to be consistent with the solvent-exfoliated materials and indicates that emulsion formation is indeed dictated by the nanosheet surface energy.

Given the robustness of emulsification to residual surfactant, the influence of pH on emulsification was investigated. Functional groups present at the edges of pristine nanosheets would not be dissociated in disperse (to reduce polarity and improve surface energy matching) but these could be deprotonated at elevated pH, such as in emulsification with a basic solution. In practice, using a standard cycloketone dispersion and KOH solution, emulsions are found to form as oil-in-water as shown in Figure 4e indicating that the deprotonation induced between pH 9 and 10 is sufficient to increase the surface energy of the nanosheets above the threshold required to invert these emulsions, around 80 mJ/m$^2$. As shown this approach can be applied for graphene, MoS$_2$ and BN, suggesting some similarity in their edge functionalities, likely S–H and N–H groups respectively. This basic inversion can be performed with solvent- or surfactant-exfoliated nanosheets, presenting an all-water-based approach for emulsifying few-layer nanosheets and controlling the orientation of subsequent emulsions.

**Conclusions**

Nanosheet-stablized emulsions represent an unexplored approach for assembly of layered materials where the combination of high surface area and functional properties have much promise for applications. Here, we have developed a framework for preparation of emulsions stablized by pristine few-layer nanosheets. Graphene- or MoS$_2$-stablized water-in-cycloketone emulsions have been shown to exhibit system-scale conductivity in their as-produced liquid form. Conductivities of ~$10^{-4}$ S/m at nanosheet volume fractions of ~$10^{-5}$ have been obtained, which represent the lowest loading level nanosheet-containing conductive composites ever reported. Their potential as emulsion inks is highlighted by the ability to drop-cast by hand into films with conductivities equivalent to other deposition techniques, facilitated by their high concentration and drying dynamics, providing spatial control, which would not be possible with standard dispersions.

To exploit the full potential of these emulsion structures, other compositions will be required to form polymer composites, charge separation interfaces, phase change materials, etc. For such applications, it will often be necessary to form oil-in-water emulsions where the water phase can be removed to form dry or solid structures. The orientation and stability of nanosheet-stablized emulsions can be understood in terms of the surface energies of the constituent phases and the inversion used to measure nanosheet surface energy to allow subsequent emulsion design. The use of basic conditions to promote deprotonation of nanosheet edge functionalities has been identified as an alternative approach to increase the surface energy of pristine nanosheets sufficiently to yield o/w emulsions. These results emphasise the robustness of the framework developed here to understand and design functional nanosheet-stablized emulsions and highlight their potential for a wide range of applications.

**Experimental Section**

*Exfoliation and emulsification*: MoS$_2$ and BN powders were purchased from Sigma Aldrich. Graphite powder was provided by Zenyatta Ventures Ltd. MoS$_2$ was subjected to an initial sonication-centrifugation step to remove impurities and very small nanosheets; the bulk powder was added to 30 mL of cyclopentanone (CPO) at an initial concentration of 25 g/L and sonicated using a Sonic Vibra-cell VCX130 at 60% amplitude for 1 hour under ice bath cooling. The dispersion was centrifuged (Thermo Scientific Sorvall Legend X1 with High Conic II rotor) at 5000 *g* for 5 mins, the supernatant containing the impurities and very small nanosheets was the discarded and the sediment was redispersed into 30 mL of fresh CPO. Graphite and BN powders were added to 30 mL of cyclohexanone at an initial concentration of 25 g/L. The subsequent sonication step used was the same for MoS2, graphite and BN; sonication using a Sonic Vibra-cell VCX130 at 60% amplitude for 3 hours under ice bath cooling. MoS2 dispersions were centrifuged at 5000 *g* for 5 mins and graphene and BN dispersions were centrifuged at 5000 *g* for 30 mins. This typically yields dispersions of nanosheets with N<10 for all materials, as confirmed with spectroscopic metrics by UV-visible extinction spectroscopy (Shimadzu UV3600Plus spectrometer). Extinction spectroscopy was also used in conjunction with previously measured extinction coefficients to determine concentration of these dispersions. Concentrations for these processing conditions are typically ~0.1 g/L. These cycloketone dispersions can be emulsified with deionised water by transferring to silanised vials and adding water at ~1:10 by volume followed by vigourously shaking by hand to homogenise. This gives nanosheet-stablized water droplets which sediment through the cycloketone continuous phase. These droplets were collected and deposited on PET to perform statistical measurements of average droplet diameter by optical microscopy (Olympus BX53-M optical microscope). In order to measure droplet size as a function of nanosheet volume fraction, the stock dispersion were diluted with cycloketone and fixed volume was emulsified with fixed volume of water to control droplet size while maintaining a fixed volume of droplets. These samples were transferred into channels milled into PTFE with copper tape contacts to allow electrical measurements using a Keithley 2600 sourcemeter. I-V characteristics were obtained and resistances normalised to channel dimensions to calculate conductivity.

*Emulsion inks:* Water-in-cycloketone emulsions of graphene and MoS$_2$ were prepared as described above. Samples were deposited by onto PET substrate heated to 80 °C by manual drop casting of 0.1 mL (per pass) of densely-packed emulsion over an area of 1 cm$^2$. The sheet resistance was measured using a Keithley 2600 sourcemeter after every deposition pass. Once dry, another 0.1 mL was deposited and this was repeated until optical microscopy showed the films to have nearly complete area coverage, around 5 passes. At this stage, AFM was performed using a Bruker Dimension Icon with ScanAsyst-Air probes to measure topography and determine approximate thickness per pass. For Raman mapping of deposited droplets, samples were deposited onto silicon wafers and their Raman spectra were mapped using a Renishaw inVia Raman microscope with 660 nm excitation using a x50 objective. The deposition process was repeated until the sheet resistance began to decrease with the reciprocal of pass number, indicating that the thickness-independent bulk-like conductivity regime had been reached.

*Solvent transfer and emulsion inversion*: In order to prepare emulsions stablized by well-exfoliated nanosheets in solvents which are conventionally considered poor for LPE, cycloketone dispersions were subjected to further centrifugation of 10000 *g* for 16 hours to result in sedimentation of almost all of the dispersed nanosheets. The cycloketone supernatant was discarded and the sediment redispersed into a new oil phase such as pentane, hexane, ethyl acetate, methyl methacrylate, dichloromethane or styrene. These oil phases span the range of surface energies of water-immiscible organic solvents and are immiscible with alternative high surface energy water phases; ethylene glycol and formamide (with the exception of ethyl acetate-formamide). As such, these combinations were used to identify emulsion orientation and stability. The solvent-transferred dispersions were emulsified with ethylene glycol, formamide and water at 1:1 by volume (to ensure sufficient oil and water phase to stablize either orientation of emulsion) and their orientation determined by identifying buoyancy and/or stability on glass or silanised vials or at the air interface. These orientations were used to verify the surface energy model presented and found to be identical for graphene, $MoS_2$ and BN emulsions whether exfoliated or bulk material was used. In order to perform the inversion experiment, a CHO dispersion was diluted to varying volume fractions of pentane and the mixed solvent dispersion emulsified with water and orientation determined. Samples between which the emulsion orientation inverted were used to calculate a range for the surface energy of the nanosheet films.

*Emulsification by surfactant-exfoliated nanosheets and basic inversion*. For the emulsification of surfactant-exfoliated nanosheets, dispersions were prepared using the exfoliation parameters described above on dispersions of graphene, $MoS_2$ or BN in 0.25 g/L aqueous Triton X-100 solution, which yields a dispersion with the minimal amount of surfactant, likely bound to the sheets rather than free in dispersion. Surfactant concentration of 0.1 g/L was found to result in significantly reduced concentration, while dispersions produced by exfoliation at higher surfactant concentration required washing by vacuum filtration and redispersion in order to allow stable emulsification. For the emulsion inversion by basic deprotonation, cycloketone dispersions were prepared and emulsified with pH 13 KOH solution, diluted to yield water phases with controlled pH, resulting in formation of buoyant oil droplets in a continuous phase of the basic solution above pH 9. Surfactant exfoliation and basic inversion can also be achieved by blending aqueous surfactant dispersions of nanosheets with KOH solution followed by emulsification with an arbitrary oil phase.

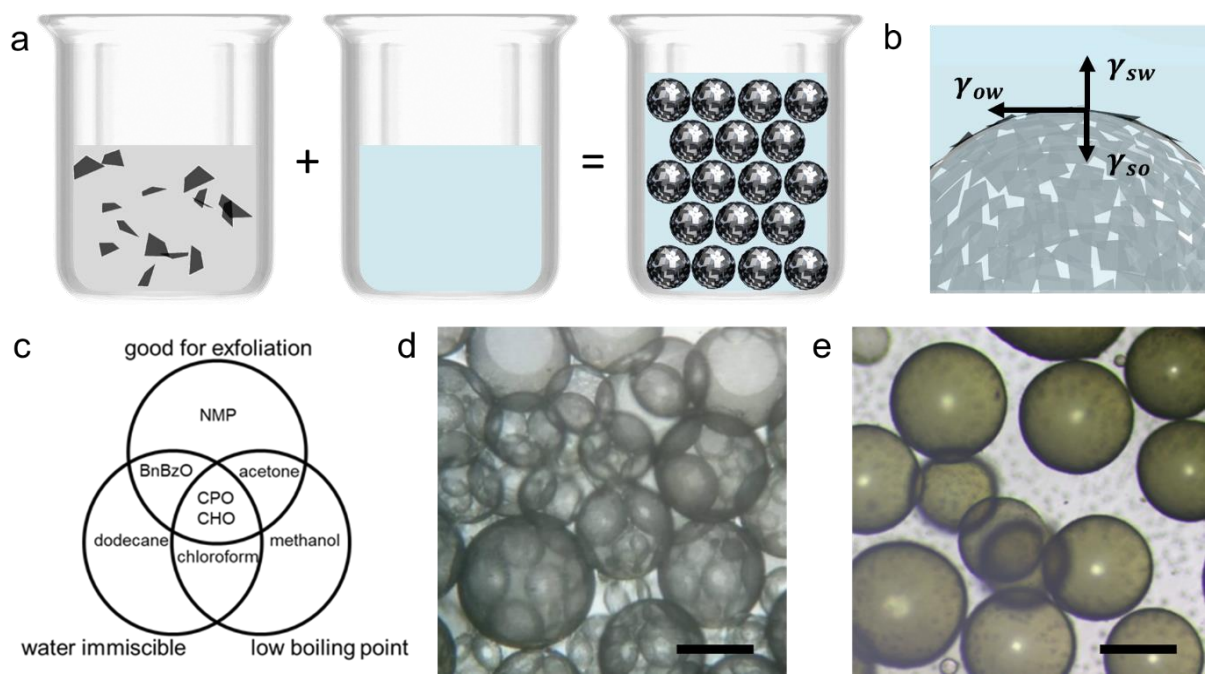

Figure 1. (a) Schematic diagram of emulsification process where nanosheets in water-immiscible solvent are homogenised with water to give water-in-oil emulsion and illustration of nanosheets on surface of a droplet. (b) Interfacial energies at three-phase boundary, which dictate emulsion stability and orientation. (c) Venn diagram illustrating solvent selection considerations for nanosheet-stablized emulsions. (d) and (e) optical micrographs of water-in-cycloketone droplets stablized by graphene and MoS$_2$ respectively, scale bar 100 μm.

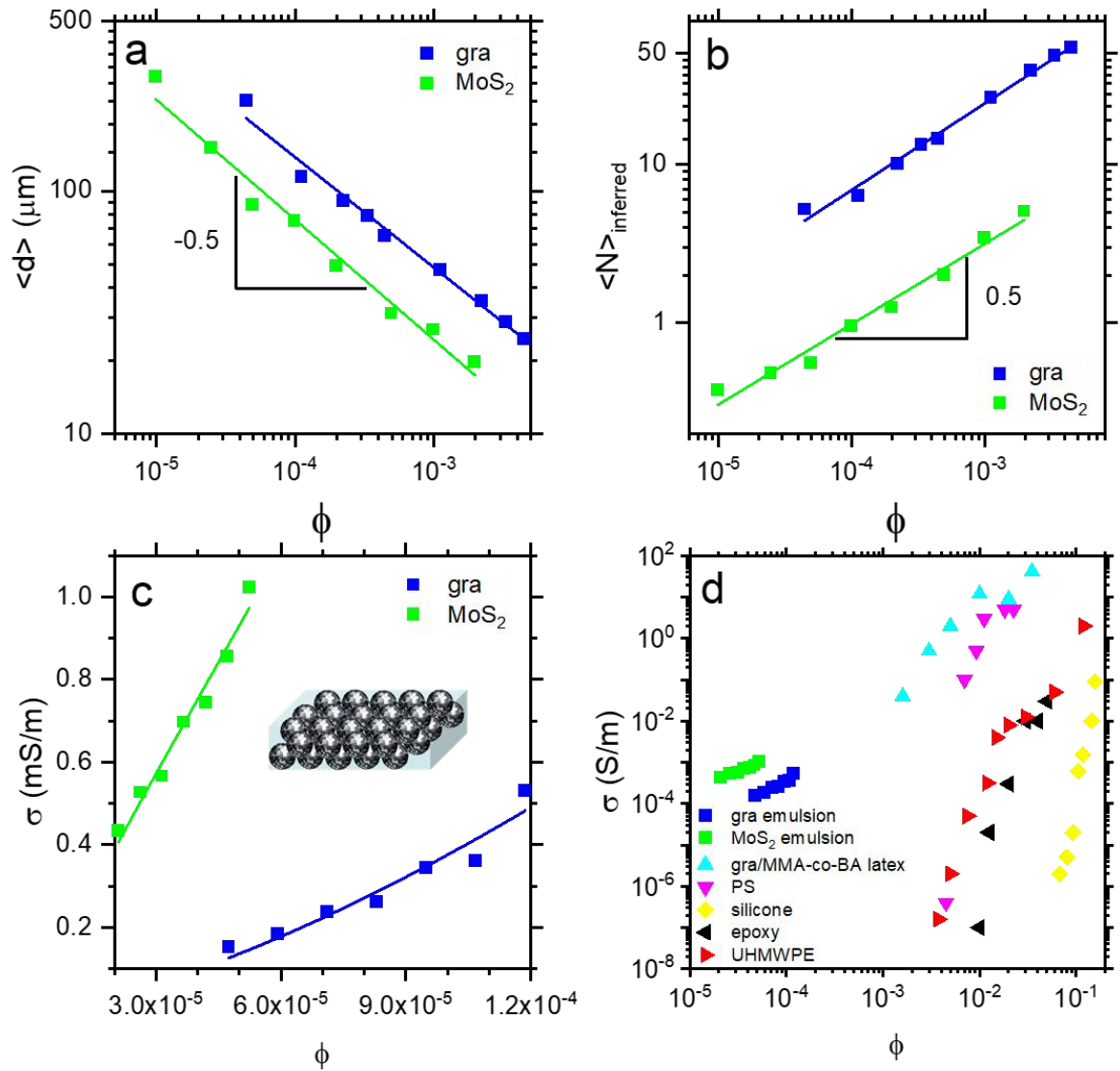

Figure 2. (a) Droplet diameter as a function of nanosheet volume fraction for graphene and MoS2 emulsions showing $\phi^{-0.5}$ dependence, attributed to $\langle N \rangle$ increasing with $\phi$ (b) Layer number, inferred from geometric model, as a function of nanosheet volume fraction with corresponding $\phi^{-0.5}$ scaling. (c) Conductivity of liquid emulsions as a function of nanosheet volume fraction. (d) Conductivity-volume fraction comparison to pristine graphene composites from the literature[5], highlight the appreciable conductivity at ultra-low loading level in the nanosheet-stablized emulsions.

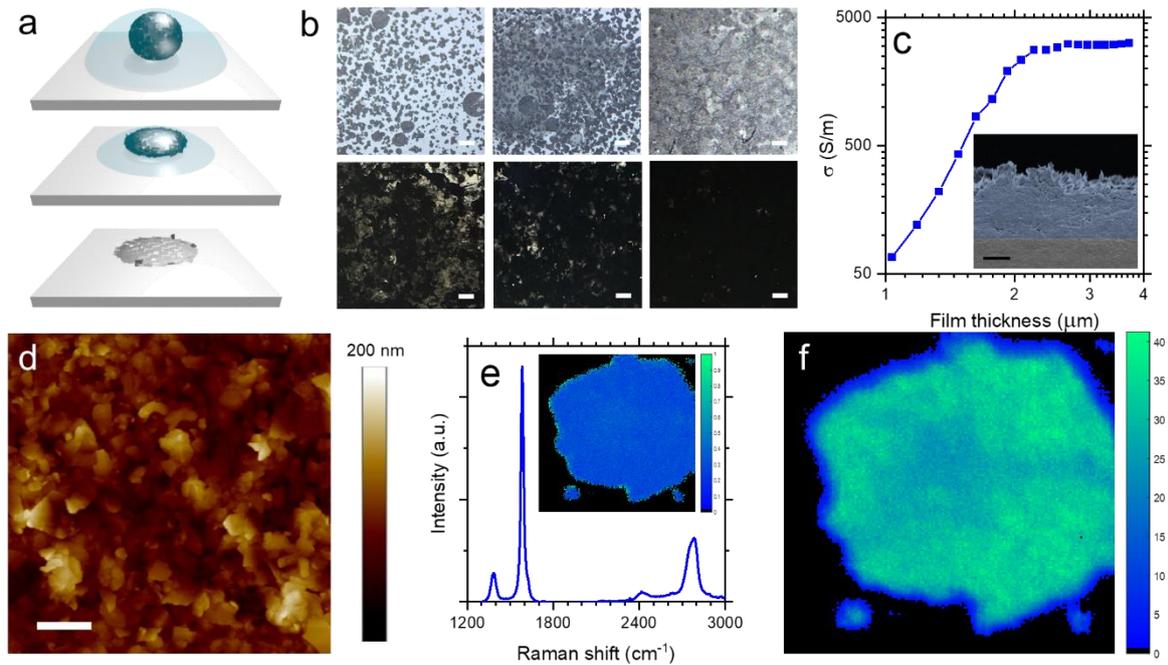

Figure 3. (a) Schematic illustration of emulsion droplet deposition, drying and collapse. (b) Low magnification optical micrographs of deposited droplets on PET showing eventual areal percolation and formation of densely-packed films. (c) Conductivity of graphene film deposited from emulsion as a function of film thickness, showing scaling attributed to deposition uniformity, which reaches expected bulk-like value. Inset: Scanning electron micrograph of film cross section (false coloured), showing dense-packed nanosheet network, scale bar 1 µm. (d) Atomic force micrograph of nanosheet film confirming dense and uniform areal packing of nanosheets deposited from a single emulsion droplet, scale bar 500 nm. (e) Raman spectrum of deposited droplet, characteristic of few-layer graphene. Inset: 2D/G ratio mapped over droplet, 30 x 30 µm image. (f) Raman map of G peak intensity illustrating uniformity of deposited film, 30 x 30 µm image.

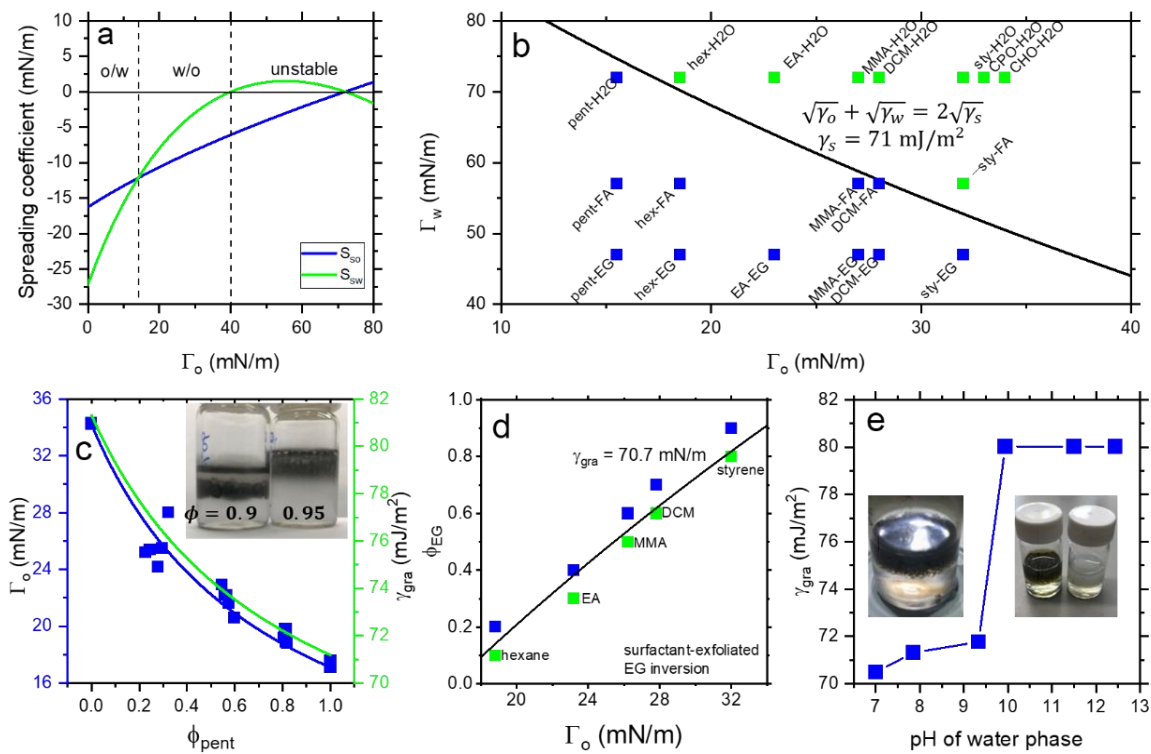

Figure 4. (a) Spreading coefficients for emulsions of graphene and water as a function of oil phase surface tension. (b) Surface tension phase diagram showing different compositions giving rise to o/w (blue) and w/o (green) emulsions which is well described by equation (6) with a surface energy of ~71 mJ/m² for all pristine nanosheets studied here. (c) Surface tension of oil and surface energy of graphene as function of pentane volume fraction as inversion experiment to determine surface energy, giving a value in good agreement with above measurement. (d) Volume fraction of ethylene glycol required for inversion as a function of oil phase surface tension for washed surfactant-exfoliated graphene, indicating that stabilisation is still dictated by the nanosheets. (e) Nanosheet surface energy as a function of pH of water phase, determined by pentane/CHO inversion. Inset: photograph of buoyant cycloketone droplets in water continuous phase, inverted at elevated pH, shown for graphene (left) and MoS$_2$ and BN (right).

**Supporting Information**

**Nanosheet-stabilized emulsions: ultra-low loading segregated networks and surface energy determination of pristine few-layer 2D materials**

*Sean P. Ogilvie\*, Matthew J. Large, Adam J. Cass, Aline Amorim Graf, Anne C. Sehnal, Marcus A. O'Mara, Peter J. Lynch, Jonathan P. Salvage, Marco Alfonso, Philippe Poulin, Alice A. K. King and Alan B. Dalton\**

**Droplet size model**

Droplet size can be related to nanosheet volume fraction (relative to the droplet phase) by equating the surface area of the droplets to that of the nanosheets. The surface area $A$ of a droplet can be related to its diameter $\langle d \rangle$ and the mass $m$, specific surface area $SSA$ and thickness as a number of monolayers of the nanosheets $\langle N \rangle$ as

$$A = \pi \langle d \rangle^2 = \frac{m \, SSA}{\langle N \rangle} \tag{1.1}$$

The mass of stabilising nanosheets can be related to their volume fraction by

$$m = \phi \rho_{2D} \pi \langle d \rangle^3 / 6 \tag{1.2}$$

By combining the above

$$\pi \langle d \rangle^2 = \frac{\phi \rho_{2D} \pi \langle d \rangle^3 \, SSA}{6 \langle N \rangle} \tag{1.3}$$

Noting that for layered materials the density and specific surface area can be related to the interlayer spacing as $c_{2D} = 1/\rho_{2D} SSA$, the above can be simplified to give a simple expression relating droplet diameter to nanosheet volume fraction

$$\langle d \rangle = \frac{6 c_{2D} \langle N \rangle}{\phi} \tag{1.4}$$

**Droplet conductivity model**

It is possible to develop a simple model for the resistor network of the system and its variation with droplet size which is in turn a function of volume fraction. A network of emulsion droplets can be approximated by resistors between droplets ($R_j$) connected by two resistors in parallel corresponding to droplet surface ($R_s$) and through-droplet ($R_d$) conductivity as shown in Fig. S1.

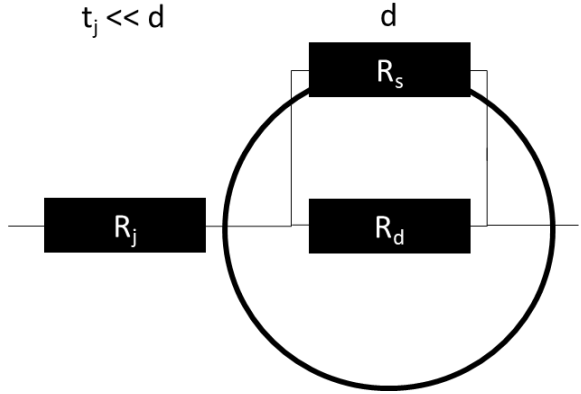

Figure S1: Unit cell of droplet network in simple conductivity model.

A two-dimensional projection of this "unit cell" as a square of side length $d$, with junction and surface thicknesses $R_j$ and $R_t$, allows calculations of the total resistance and normalisation of the unit cell geometry.

$$R_t = R_j + \frac{1}{(\frac{1}{R_s} + \frac{1}{R_d})} \tag{2.1}$$

$$R_t = R_j + \frac{R_s R_d}{(R_d + R_s)} \tag{2.2}$$

This total unit cell resistance can be related to the conductivity and dimensions of the constituent phases and subsequently equated to the conductivity and dimensions of the unit cell itself as

$$R_t = \frac{t_j}{\sigma_j d^2} + \frac{\frac{1}{\sigma_s t_s} \cdot \frac{1}{\sigma_d d}}{(\frac{1}{\sigma_d d} + \frac{1}{\sigma_s t_s})} = \frac{t_j}{\sigma_j d^2} + \frac{1}{\sigma_s t_s + \sigma_d d} \equiv \frac{1}{\sigma d} \tag{2.3}$$

$$\sigma = \left[ d(\frac{t_j}{\sigma_j d^2} + \frac{1}{\sigma_s t_s + \sigma_d d}) \right]^{-1} \tag{2.4}$$

$$\sigma = \frac{1}{d}(\frac{t_j}{\sigma_j d^2} + \frac{1}{\sigma_s t_s + \sigma_d d})^{-1} \tag{2.5}$$

Depending on whether the conduction is dominated by the droplets and surfaces or the junctions, this model will be dominated by the former or latter terms respectively. Where the droplets are much more conductive than the junctions, such as for water droplets stablized by thick conductive nanosheet films in a very insulating oil phase, this leads to a decreasing conductivity with increasing loading, as observed in our previous work[1]

$$\frac{1}{\sigma_s t_s + \sigma_d d} \to 0 \tag{2.6}$$

$$\sigma = \frac{1}{d}\frac{\sigma_j d^2}{t_j} = \frac{\sigma_j d}{t_j} \tag{2.7}$$

By contrast, for coalesced emulsion polymer composites, where any inter-droplet junction resistances are reduced, this model simplifies to give a linear increase in conductivity with loading level

$$\frac{t_j}{\sigma_j d^2} \to 0 \tag{2.8}$$

$$\sigma = \frac{\sigma_s t_s + \sigma_d d}{d} \tag{2.9}$$

$$\sigma = \frac{\sigma_s t_s}{d} + \sigma_d \tag{2.10}$$

$$\sigma = \frac{\sigma_s}{6}\phi + \sigma_d \tag{2.11}$$

In practise, the all-liquid emulsion networks studied in this manuscript exhibit some intermediate behaviour which can be fitted to the original model but is also functionally equivalent to a power law in the range studied, as shown in Fig. 2c.

**Non-Newtonian rheology**

Proof-of-concept rheological measurements were performed to demonstrate the non-Newtonian behaviour of these nanosheet-stablized emulsions. Shear-rate dependent shear stress and viscosity are shown for a representative graphene-stablized water-in-CHO emulsion in Fig. S2 exhibit the expected shear thinning behaviour with a characteristic power law scaling $\eta \approx 0.1\,\dot{\gamma}^{-0.58}$ and little hysteresis.

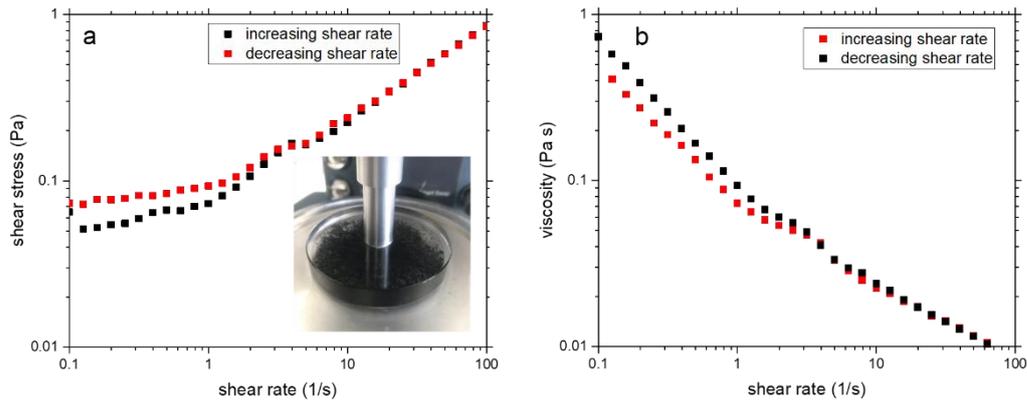

Figure S2: (a) shear stress as a function of shear rate during parallel plate rheology of representative graphene-stablized water-in-CHO emulsion. Inset: photograph of experimental setup (b) viscosity as a function of shear rate.

In order to be suitable for inkjet deposition, these emulsions also require a viscosity (~0.01 Pa.s) greater than that of common solvents at the shear rates applied during jetting (~$10^6$ s$^{-1}$). While these water-in-cycloketone emulsions reach the viscosity required for inkjet printing at 100 s$^{-1}$, $10^4$ times lower shear rate than during jetting, it it possible that viscosity will rapidly saturate at higher shear rates as shown previously for clay-stablized water-in-oil emulsions[2]. Alternatively, it may be possible to use dilute emulsions (with lower ratio of droplet to continuous phase) which are known to exhibit Newtonian behaviour with viscosity independent of shear rate[3] to ensure the desired viscosity during jetting. However, this does mean reducing the concentration of the emulsion ink and potentially using a high viscosity (likely high boiling point) continuous phase, the selection of which must also satisfy other criteria for surface energy, nanosheet dispersability, etc. A more practical alternative might be to manipulate the shear rate-dependent viscosity by controlling emulsion droplet size. It is well known that smaller droplets in a concentrated emulsion give rise to increased viscosity[3,4] which presents a route to ensure sufficient viscosity during inkjet.

**Surface energy model**

The stability and orientation of solid-stablized emulsions can be related to the spreading coefficients and constituent interfacial energies. The spreading coefficients for the solid-oil and solid-water are given by

$$S_{so} = \gamma_{so} - \gamma_{sw} - \gamma_{ow} \qquad (4.1)$$

$$S_{sw} = \gamma_{sw} - \gamma_{so} - \gamma_{ow} \qquad (4.2)$$

Where the $\gamma_{so}, \gamma_{sw}$ and $\gamma_{ow}$ are the interfacial energies at the solid-oil, solid-water and oil-water interfaces. The above definitions can be combined to give

$$S_{so} + S_{sw} = -2\gamma_{wo} \tag{4.3}$$

Since interfacial tensions/energies are positive, spreading coefficients can only have the same sign (and thereby form a stable emulsion) if that sign is negative. If both spreading coefficients are negative, the stability criteria can be expressed as

$$\gamma_{so} - \gamma_{sw} < \gamma_{wo} \tag{4.4}$$

$$\gamma_{sw} - \gamma_{so} < \gamma_{wo} \tag{4.5}$$

Since $S_{so} - S_{sw} = -(S_{sw} - S_{so})$, one of the above equations will always be satisfied and the criterion reduces to

$$|\gamma_{so} - \gamma_{sw}| < \gamma_{ow} \tag{4.6}$$

Based on the geometric and harmonic mean models, it can be intuitively argued that it is most easily satisfied by $\gamma_o \ll \gamma_w$ (giving large $\gamma_{ow}$) and $\gamma_s \approx \gamma_o$ and $\gamma_s \approx \gamma_w$ (giving $\gamma_{so} \approx \gamma_{sw}$) and the difference is small), which requires that $\gamma_o < \gamma_s < \gamma_w$, as illustrated in Fig. 4a, although this is more challenging to demonstrate rigourously.

However, in order to explicitly state this condition, interfacial energy models are required. The orientation of an emulsion (o/w or w/o) is also determined by the spreading coefficients, i.e. whichever is more negative forms the droplet phase; o/w for $S_{so} < S_{sw}$ and w/o for $S_{so} > S_{sw}$. As such, the point at which they are equal can be considered the inversion threshold for an emulsion. This can be simplified (by definition and without any empirical models) as

$$\gamma_{so} = \gamma_{sw} \tag{4.7}$$

Subsequently, simple models for interfacial energies can be substituted such as[5]

$$\gamma_{ab} = \gamma_a + \gamma_b - 2\sqrt{\gamma_a \gamma_b} \tag{4.8}$$

$$\gamma_{ab} = \gamma_a + \gamma_b - 4\frac{\gamma_a \gamma_b}{\gamma_a + \gamma_b} \tag{4.9}$$

Incorporating the geometric mean model (Equation 4.8) into Equation 4.7 gives an expression which describes the inversion threshold of emulsions as a function of the constituent surface energies

$$\gamma_s + \gamma_o - 2\sqrt{\gamma_s \gamma_o} = \gamma_s + \gamma_w - 2\sqrt{\gamma_s \gamma_w} \tag{4.10}$$

$$\gamma_o - 2\sqrt{\gamma_s \gamma_w} = \gamma_w - 2\sqrt{\gamma_s \gamma_w} \tag{4.11}$$

Substituting $\gamma_o = x^2$ and $\gamma_w = y^2$

$$x^2 - 2\sqrt{\gamma_s}x = y^2 - 2\sqrt{\gamma_s}y \tag{4.12}$$

$$x^2 - y^2 = 2\sqrt{\gamma_s}x - 2\sqrt{\gamma_s}y \tag{4.13}$$

$$(x-y)(x+y) = 2\sqrt{\gamma_s}(x-y) \tag{4.14}$$

Cancelling $(x - y)$ gives

$$x + y = 2\sqrt{\gamma_s} \tag{4.15}$$

Finally, re-expressing in terms of surface energies yields

$$\sqrt{\gamma_o} + \sqrt{\gamma_w} = 2\sqrt{\gamma_s} \tag{4.16}$$